\documentclass[aps,prb,twocolumn,superscriptaddress,floatfix,showpacs]{revtex4-1}

\usepackage[T1]{fontenc}
\usepackage[english]{babel}

\usepackage{amssymb,amsmath}
\usepackage{graphicx}
\usepackage{color}

\def\rmd{\mathrm{d}}
\def\rmd{\mathrm{d}}
\def\rmc{\mathrm{c}}
\def\rms{\mathrm{s}}
\def\rmgs{\mathrm{GS}}
\def\rmb{\mathrm{B}}
\def\rmf{\mathrm{F}}
\def\up{\uparrow}
\def\down{\downarrow}
\def\Jaf{J_{\mathrm{\tiny ex}}}
\def\Sf{S_f}
\def\Tf{T_f}

\def \plotwi{8.3cm}
\def \plotwiless{7cm}

\newcommand{\bonn}{HISKP, University of Bonn, Nussallee 14-16, D-53115 Bonn, Germany.}
\newcommand{\dpmc}{DPMC-MaNEP, University of Geneva, 24 Quai Ernest-Ansermet, 1211 Geneva 4, Switzerland.}
\newcommand{\dpt}{DPT, University of Geneva, 24 Quai Ernest-Ansermet, 1211 Geneva 4, Switzerland.}

\begin{document}
\title{Competition of spin and charge excitations in the Hubbard model}
\author{Bruno Sciolla}
\affiliation{\bonn}
\author{Akiyuki Tokuno}
\author{ Shun Uchino}
\affiliation{\dpmc}
\author{ Peter Barmettler}
\affiliation{\dpt}
\author{ Thierry Giamarchi}
\affiliation{\dpmc}
\author{ Corinna Kollath}
\affiliation{\bonn}


\begin{abstract}
Motivated by recent experiments with ultracold fermionic atoms in optical lattices, we study finite temperature magnetic
 correlations, as singlet and triplet correlations, and the double
 occupancy  in the one-dimensional Hubbard model.
We point out that for intermediate interaction strengths the double occupancy has an intriguing doubly non-monotonic temperature dependence due to the competition between spin and charge modes, related to the Pomeranchuk effect.
Furthermore, we determine properties of magnetic correlations in the temperature regimes relevant for current cold atom experiments and discuss effects of the trap on spatially integrated observables. We estimate the entropy and the temperature reached in the experiment by Greif {\it et al.}, Science {\bf 340}, 1307 (2013).
\end{abstract}
\maketitle

\section{Introduction}
\label{sec:introduction}

Understanding the effects of strong correlations is one of the most challenging
problems in quantum physics. Such effects have besides its intrinsic fundamental
interest far reaching consequences on material science and condensed matter physics
since in many materials strong correlations occur. Among the strongly correlated materials some typical examples include many oxydes and Mott insulators \cite{imada98_mit_review}, 
nanotubes \cite{dresselhaus_book_fullerenes_nanotubes}, organic materials \cite{lebed_book_1d} and most probably the high temperature superconductors \cite{bonn_hightc_review}.

The effects of strong correlations can best be studied using simplified models, the most celebrated one is the so-called Hubbard model~\cite{hubbard63model}. The Hubbard model includes the competition of a tight-binding kinetic term of amplitude $t$ and a local on-site interaction with amplitude $U$.

Despite important efforts, many of properties of the Hubbard model are not yet fully understood even in one and infinite dimension, where in principle it can be solved exactly~\cite{giamarchi_book_1d,georges_d_infini}. Especially in two dimensions its solution is still largely elusive. For a filling of one particle per site
however, it is well-known that there are insulating phases dominated by interactions (Mott insulators) and
 antiferromagnetic phases, due to the existence of the superexchange mechanism~\cite{Georges.Giamarchi/LesHouches2012}. 
Away from half-filling the game's afoot, given the difficulty to tackle this model analytically or numerically.

Recently, cold atoms have provided a very nice realization of the Hubbard model, and thus
offered the possibility to act as quantum simulators to study the physics of this problem.
The Hubbard model is realized with fermionic neutral atoms trapped in an optical lattice~\cite{bloch_cold_lattice_review}. Different internal states of the atoms are used to represent the spin degrees of freedom. The interactions between these different states, the van der Waals interaction, is short ranged and can be tuned via a Feshbach resonance. The whole setup allows remarkable control on the various physical parameters.
The realizations of the Hubbard like models in optical lattices allowed for the direct observation of a Mott insulating phase for bosonic \cite{Greiner.etal/Nature415.2002} and fermionic atoms \cite{Jordens2008,Schneider/2008}.
Observing more complex quantum phases as the antiferromagnetic order or even unconventional superconducting phases has proven quite difficult given the high temperatures present in the atomic gases of the order of $~0.1 E_F$ where $E_F$ is the Fermi energy. It has thus become a holy grail to observe quantum magnetism in such systems. Coherent magnetism has previously been observed in bosonic
systems, for a two site problem~\cite{Trotzky2008} and
for hard core bosons
mimicking an XY model~\cite{Struck.etal/Science333.2011}
or bosons in a tilted lattice mimicking a transverse field Ising model~\cite{Simon.etal/Nature472.2011}.
The coherent propagation of a magnetic excitations has been measured
in an experimental realization of the ferromagnetic Heisenberg model~\cite{Fukuhara.etal/NatPhys9.2013}, 
but the antiferromagnetism was still out of reach.
The first evidence of short range antiferromagnetism in fermionic systems has been reported in the Hubbard model using the modulation spectroscopy of the optical lattice \cite{GreifEsslinger2011} 
as proposed in Ref.~\cite{KollathGiamarchi2006}. A more direct evidence has very recently observed in
anisotropic fermionic Hubbard structures~\cite{Greif.etal/Science340.2013}. The nearest neighbor singlet and triplet correlations were found to have significant imbalance,
showing that 
the temperature was low enough to sustain detectable short
range antiferromagnetic correlations.

Motivated by the possibility to observe the short range spin correlations, we study here these correlations for the case of the one dimensional Hubbard model, as a function of the
interaction and the temperature. We compute in particular the nearest neighbor magnetic correlations such as the ratio between
singlet and triplet excitations using both a numerical procedure, i.e., time
dependent density-matrix renormalization group (DMRG), and analytical arguments.
We determine the optimal regimes of parameters to observe antiferromagnetic correlations and discuss the consequences of such an order
in terms of Pomeranchuk effect. We also take into account the presence of the trap as well as the doping of the system. We additionally estimate the temperature present in the experiments \cite{Greif.etal/Science340.2013}.

The plan of the paper is as follows. The model and its magnetic properties are introduced in Sec.~\ref{sec:model}, as well as the different methods that we use. The non-monotonic temperature dependence of double occupancy is described in Sec.~\ref{sec:double_occ}. We then introduce a phenomenological model for the low energy modes in Sec.~\ref{sec:phenomenology}, and use it to explain the origin of the double non-monotonicity as a competition between charge and spin degrees of freedom.
We then analyze the singlet and triplet correlations as a function of temperature and interactions in Sec.~\ref{sec:singlet_triplet_T}.
The effect of the trap is discussed in Sec.~\ref{sec:trap_singlet_triplet}. In the regimes that we study, the local density approximation is still valid, as seen in Sec.~\ref{sec:trap_lda}.
Finally, we extract the entropy and temperature reached in the experiment~\cite{Greif.etal/Science340.2013} in Sec.~\ref{sec:trap_exp}.
Conclusions and perspectives are presented in Sec.~\ref{sec:conclusion}.

\section{The Hubbard model}
\label{sec:model}
We consider here cold fermions confined to one-dimensional tubes along which a lattice potential is applied. This system is well described over a wide range of parameters by the single band Hubbard model:
\begin{align}
\label{eq:hamiltonian}\hat{H} = -t \sum_{ i,\sigma=\uparrow,\downarrow} \left( \hat{c}^\dagger_{i,\sigma}  \hat{c}_{i+1,\sigma} + \textrm{h.c.} \right) + U \sum_{i} \hat{n}_{i,\uparrow}\hat{n}_{i,\downarrow}
\end{align}
where $\hat{c}_{i,\sigma}^\dagger,\hat{c}_{i,\sigma}$ are the fermionic creation and annihilation operators and $\hat{n}_{i,\sigma} = \hat{c}_{i,\sigma}^\dagger\hat{c}_{i,\sigma}$ the density operator. The different spin states represent typically different hyperfine states of the fermionic atoms.
$t$ is the tunneling amplitude and $U>0$ is the repulsive on-site interaction between different spin species and in many setups can be tuned experimentally using a Feshbach resonances. 

This model comprises a lot of interesting physics such as Mott insulating
and liquid phases~\cite{gebhard/1997,giamarchi_book_1d,Essler.etal/book.2010}.
A particularity of the one-dimensional model is that a low energy single particle excitation separates rapidly into charge and spin mode~\cite{gebhard/1997, giamarchi_book_1d,Kollath/2005}. In the following, the charge density is defined as the total atom density operator $\hat{n}_\up +\hat{n}_\down$, whereas the spin density operator is the difference of the different spin states $\hat{S}^z = (\hat{n}_\up - \hat{n}_\down)/2$.

Away from half filling both modes are gapless leading to a liquid ground state. In contrast, at half-filling, any finite repulsive interaction $U>0$ causes a charge gap and the ground state is a Mott insulator.
For small $U/t \ll 1$, the charge gap is exponentially small $\propto e^{-t/U}$, whereas in the large $U/t$ limit it is approximately proportional to $U$. Due to the presence of the gap, charge fluctuations are strongly suppressed below temperatures of about a tenth of the gap~\cite{Shiba1972,Usuki1990}.

As for the low energy magnetic properties, in the regime of large interactions $U \gg t$, the model can be mapped onto a Heisenberg chain~\cite{Georges.Giamarchi/LesHouches2012}. The important parameter is the antiferromagnetic exchange coupling $ \Jaf=4 t^2/U$. Intuitively this is explained by a gain in energy due to a virtual hopping process when neighboring sites are occupied by different spin species. The resulting antiferromagnetic coupling leads to enhanced antiferromagnetic spin correlations below $k_BT<\Jaf$ at large interaction strength. At zero temperature, the correlations algebraically decay and their asymptotics can be well described within the Tomonaga-Luttinger (TL) liquid description~\cite{giamarchi_book_1d}, in which the low energy spin modes have a gapless linear dispersion.

\begin{figure}
\includegraphics[width=\plotwi]{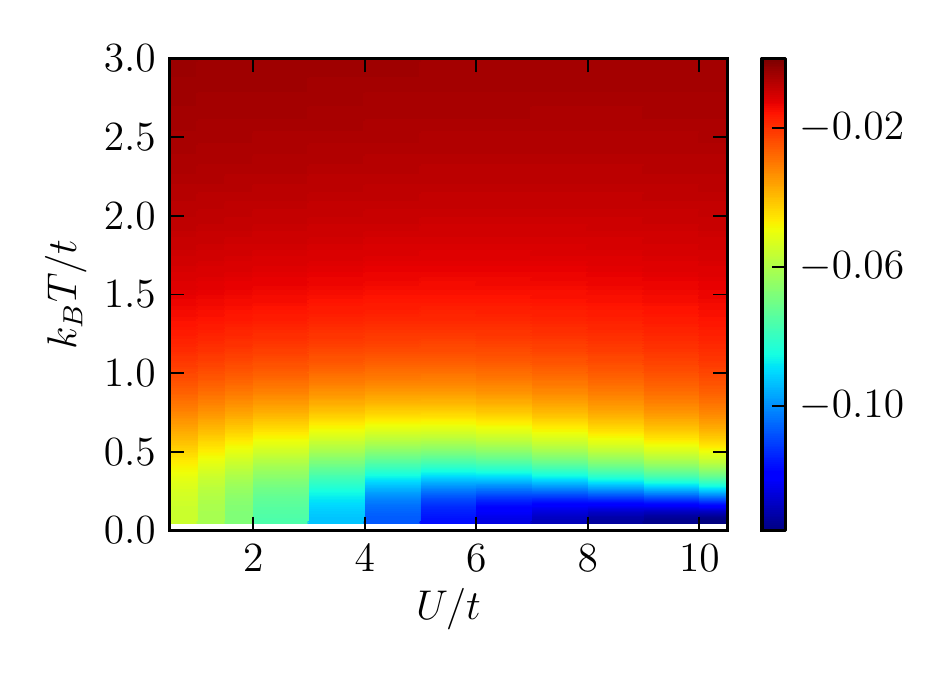}\\
\caption{\label{fig:colorplota} (Color online). Nearest-neighbor spin
 correlations
 $\langle \hat{S}^z_i \hat{S}^z_{i+1} \rangle$ as a function of $U/t$
 and $k_B T/t$, from DMRG. One can notice that the spin
 correlations are stronger within the intermediate $U/t$
 region at finite temperatures. For large $U/t$, the spin correlations decay on the
 scale of the antiferromagnetic exchange
 $\Jaf = 4 t^2/U$.}
\end{figure}

\begin{figure}
\includegraphics[width=\plotwiless]{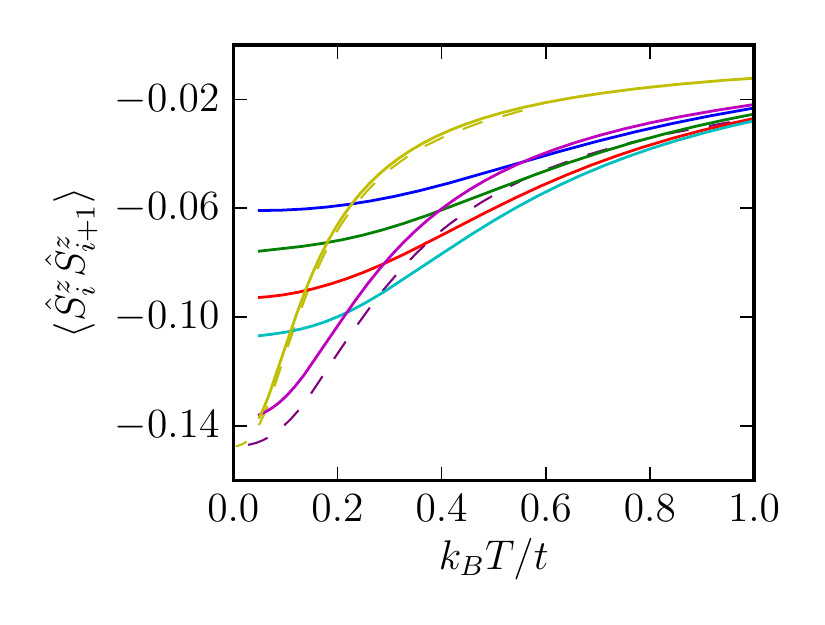}\\
\caption{\label{fig:spin_cor}(Color online). Nearest-neighbor spin correlations $\langle \hat{S}^z_i \hat{S}^z_{i+1} \rangle$ versus $k_B T/t$. Full lines: DMRG result for $U/t = 1,2,3,5,10, 20$ from top to bottom increasing $U/t$ (for low $k_B T/t$). Dashed lines: exact correlations for the Heisenberg chain of coupling $\Jaf = 4 t^2/U$ for $U/t=10,20$. For large interactions $U/t \gtrsim 10$, the spin correlations decay on the scale of the antiferromagnetic exchange $\Jaf = 4 t^2/U$. The agreement with correlations in the Heisenberg chain is excellent for $U/t \gtrsim 20$.}
\end{figure}

The temperature dependence of nearest-neighbor spin correlations
$\langle \hat{S}^z_i \hat{S}^z_{i+1} \rangle$ are shown in Figs.~\ref{fig:colorplota} and~\ref{fig:spin_cor}. Even for these short-range correlations, the lines of constant correlations of Fig.~\ref{fig:colorplota} have the typical dome-like structure at low temperatures, known from the three dimensional phase diagram.
The non-monotonicity along the fixed-$T$ lines can be understood
considering the different limits. In the strongly interacting regime $U
\gg t$, the system is characterized by the spin exchange coupling $\Jaf=4 t^2/U$, which leads to a decrease of spin correlations while increasing $U$.
In contrast, at low interaction strength charge fluctuations are also present and lead to a destruction of the spin correlations above the charge gap. Thus spin correlations are maximally robust against thermal fluctuations
in the regime of intermediate interactions.
In Fig.~\ref{fig:spin_cor}, we show that the spin correlations extracted from the mapping to a Heisenberg chain of coupling $\Jaf = 4 t^2/U$ are in good agreement with the spin correlations for $U/t \gtrsim 20$.
The spin correlations at finite temperature in the Heisenberg chain have been obtained
by using the Bethe ansatz method
\cite{takahashi1971one,PhysRevLett.26.1301,koma1988one,PhysRevB.43.5788,klumper1998spin,takahashi2005thermodynamics}.

 Whereas many
different methods have been applied in order to investigate the thermodynamic and ground state properties, the calculation of correlation functions at finite temperature is more involved. For example within the Bethe ansatz ~\cite{Essler.etal/book.2010}, the free energy can be derived from the thermodynamic Bethe ansatz \cite{Takahashi1972} or the quantum transfer matrix \cite{Juttner1998},
but correlation functions are much more difficult to extract.
In addition the presence of the charge gap makes it hard to employ field
theoretical analysis straightforwardly.
This motivates us to use a variety of methods which complement
well each other: the finite temperature DMRG on the numerical side, and a phenomenological approach which combines bosonization and Bethe Ansatz on the analytical side.
In the DMRG simulation, a matrix product representation~\cite{Schollwock2011} of the finite temperature density matrix $\hat{\rho}=e^{-\beta \hat{H}}/Z$ is obtained by evolving the trivial infinite temperature state in imaginary time~\cite{Verstraete2004,Zwolak2004,White2004,Schollwock2011}.
Our implementation conserves the commutator of total magnetization and charge with the density matrix \cite{Muth2011}, the number of particles is fixed by a chemical potential.

For the imaginary time evolution we use a fifth order Suzuki-Trotter decomposition with a typical step of $\Delta \beta\approx 10^{-3} t$.
The convergence with respect to the number of retained states $M$ has been checked, with $M \leq 516$ states.
We use the periodic matrix product representation for the thermodynamic limit~\cite{Vidal2004} which is orthogonalized~\cite{Orus2008,Mcculloch2008} before evaluation of observables.
For trapped systems we use the finite size matrix product representation~\cite{Schollwock2011}.

\section{Properties at half-filling}
\label{sec:summary}
Here we analyze in detail the charge and spin fluctuations
which occur at half filling in the low-temperature regime and their
experimental signatures. With this aim, we mainly focus on
experimentally measurable quantities as the number of double occupied
sites and the nearest neighbor spin singlet and triplet correlations. We
find that the competition between charge and spin fluctuations is
most interesting in the intermediate regime of interaction strength,
where the energy scales are comparable. The competition causes a {\it double Pomeranchuk} effect, i.e. a double non-monotonicity in the behavior of the double occupancy versus temperature.

\subsection{Double occupancy}
\label{sec:double_occ}
The double occupancy $n_d = \langle \hat{n}_{i,\uparrow}\hat{n}_{i,\downarrow} \rangle$ reveals intriguing physics on a broad range of energy scales, in particular at low energies due to the interplay between spin and charge degrees of freedom. The double occupancy has been considered in previous theoretical studies in one dimension~\cite{Gorelik.etal/PRA85R.2012}, two dimensions~\cite{Paiva/104.066406} and three dimensions~\cite{Georges.etal/PRB48.7167,Werner.etal/PRL95.2005,Dare.etal/PRB76.064402, DeLeo.etal/PRA83.2011} and experimentally in three dimensions~\cite{Kohletal/2005,Jordensetal/2010}.

At high temperatures the double occupancy grows monotonically while increasing the temperature towards its infinite temperature value $n_d = 1/4$, since more and more density fluctuations are created. In contrast, it has been observed that at low temperatures the double occupancy decays with increasing temperature.
This decrease is at first sight counter-intuitive, since it lowers the charge fluctuations. However, it is understood that it facilitates spin excitations and is preferred in terms of entropy.
The entropy density $s$ and the double occupancy are related by the
(exact) Maxwell relation $\frac{ 1}{ k_B} \frac{\partial s}{\partial U} = - \frac{ 1}{ k_B}\frac{
\partial n_d}{\partial T}$. This effect is the analog of the
Pomeranchuk effect \cite{RichardsonRevModPhys.69.683,DeLeo.etal/PRA83.2011,Dare.etal/PRB76.064402,Werner.etal/PRL95.2005,Georges.etal/PRB48.7167,Paiva/104.066406}

\begin{figure}
\includegraphics[width=\plotwi]{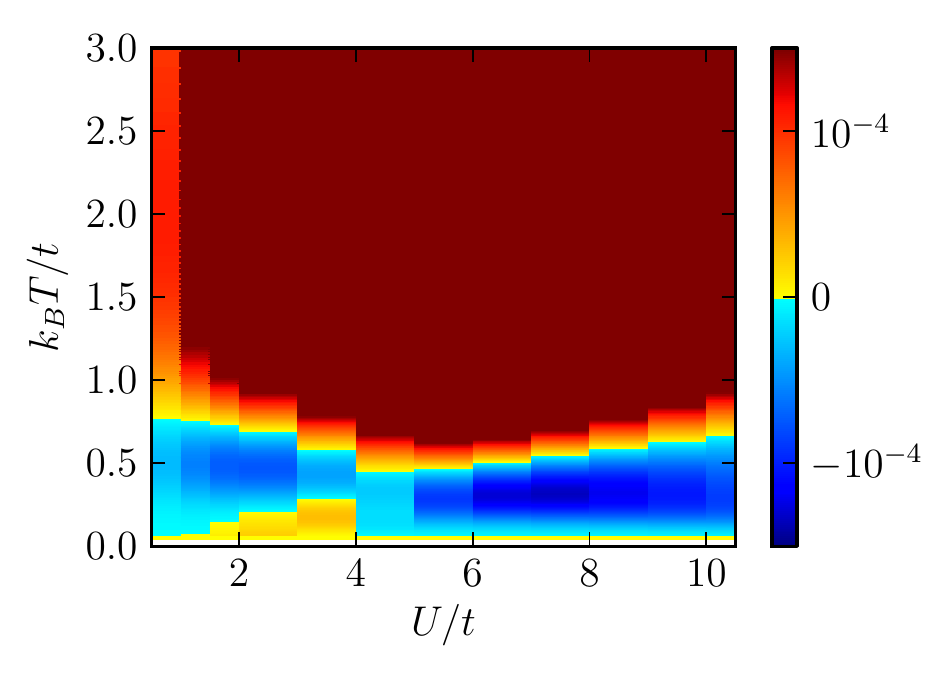}
\caption{\label{fig:colorplotb}(Color online). Temperature derivative of the double occupancy $\displaystyle \partial n_d / \partial (k_B T)$ versus $(U/t,k_B T/t)$, from DMRG. This quantity allows to characterize the regimes of growths or decays of the double occupancy with temperature. The successive decay and growth of double occupancy, for large $U/t$, is the signature of the Pomeranchuk effect. For intermediate interactions $1 \lesssim U/t \lesssim 4$, the double occupancy successively grows, decays and grows again with temperature. This signals the presence of a double Pomeranchuk effect.}
\end{figure}

\begin{figure}
\includegraphics[width=\plotwiless]{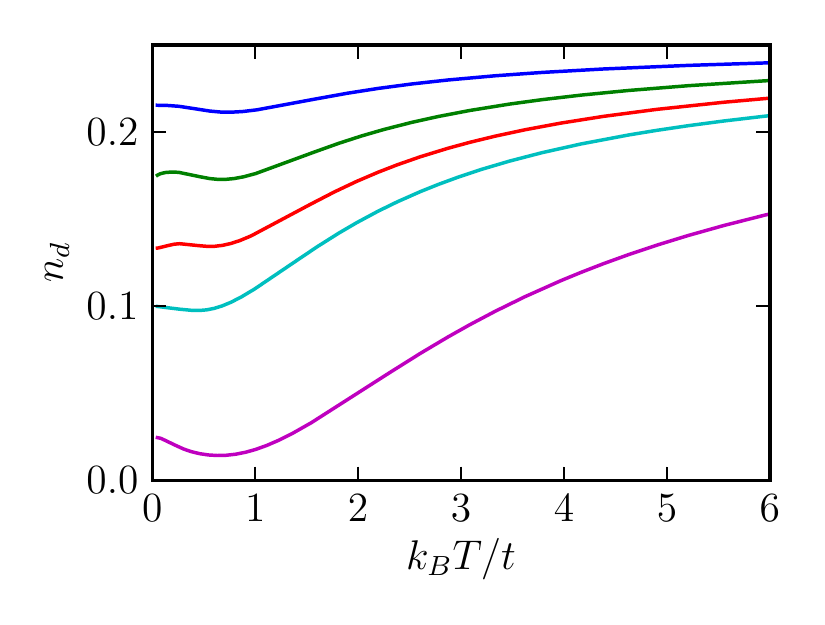}\\
\caption{\label{fig:double_occ}(Color online). Double occupancy $n_d$ versus $k_B T/t$, from DMRG, for $U/t = 1,2,3,5,10$ from top to bottom increasing $U/t$. For large $U/t$, the conventional Pomeranchuk effect is the successive decay and growth of double occupancy with growing temperature. For intermediate interactions $1 \lesssim U/t \lesssim 4$, the double occupancy successively decays (not visible here), grows, decays and grows again with temperature, which we refer to as a double Pomeranchuk effect.}
\end{figure}

We observe this phenomenon in the regime of large interactions $U \gg
t$, as shown in Fig. \ref{fig:colorplotb}, where it is shown that
$\displaystyle \partial n_d / \partial(k_B T)$ is successively negative
and positive as temperature grows. Cuts of $n_d$ at fixed $U/t$ are also
shown in Fig.~\ref{fig:double_occ}. In the regime of low temperatures
and intermediate interactions $1 \lesssim U/t \lesssim 4$, we observe,
however, that there is an additional low temperature regime of growth of
double occupancy. As we will see, this is the sign of a double Pomeranchuk effect.
According to this picture, the double occupancy actually decays (on lower temperatures than shown), grows, then decays again and finally grows at higher temperatures. In the next sections, we endeavor to understand this phenomenon in terms of a competition between charge and spin degrees of freedom using a phenomenological model.

\subsection{Phenomenological approach}
\label{sec:phenomenology}
In order to identify the physical mechanism underlying the {\it double Pomeranchuk effect} seen in the DMRG results, we discuss the finite temperature behavior of the double occupancy within an effective model.

Within this model we use the well-known fact that the low-energy excitations of the one-dimensional Hubbard model
at half-filling consists of decoupled spin and charge sectors with linear and massive dispersion, respectively~\cite{giamarchi_book_1d}. Further, we assume that the spin and charge excitations obey bosonic
and fermionic statistics, respectively.
This leads to an effective low energy Hamiltonian of the form
\begin{equation}
\label{eq:eff_H}
\hat{H}=E_{\rmgs}+\hat{H}_{\rms}+\hat{H}_{\rmc}.
\end{equation}
Here $E_{\rmgs}$ is the ground state energy of the Hubbard model.
The spin Hamiltonian is given by
\begin{equation}
 \hat{H}_{\rms}
 =\sum_{\sigma=\uparrow,\downarrow}\sum_{k}
  \epsilon_{\rms}(k)\hat{s}^{\dagger}_{\sigma}(k)\hat{s}_{\sigma}(k),
 \label{eq:Hs}
\end{equation}
where the linear dispersion $\epsilon_{\rms}(k)=\hbar v_{\rms}|k|$ corresponds to
the excitation energy of the spin sector with spin velocity $ v_{\rms}$. The operators $\hat{s}^{\dagger}_{\sigma}(k),\hat{s}_{\sigma}(k)$ are assumed to have bosonic commutation relations.

The charge part is expressed by fermionic particle $p$ and hole $h$ excitations
\begin{equation}
 \hat{H}_{\rmc}
 =\sum_{k}\epsilon_{\rmc}(k)
  \left[ \hat{p}^{\dagger}_{k} \hat{p}_{k}+\hat{h}_{k}\hat{h}^{\dagger}_{k} \right]
 \label{eq:Hc}
\end{equation}
with a massive dispersion $\epsilon_{\rmc}(k)=\sqrt{(\hbar
v_{\rmc}k)^2+\Delta_{\rmc}^2}$, with a sound velocity $v_{\rmc}$ and a gap $\Delta_{\rmc}$. The particles and holes have fermionic statistics and are related by the particle-hole symmetry. 

The introduced parameters, $E_{\rmgs}$, $v_{\rms}$, $v_{\rmc}$, and
$\Delta_{\rmc}$, depend on the hopping $t$ and the interaction $U$ and
are determined by Bethe Ansatz~\cite{Essler.etal/book.2010} as follows:
\begin{align}
 E_{\rmgs}
 &=-4t \int_{0}^{\infty}\!\!\frac{d\omega}{\omega}\,
     \frac{J_{0}(\omega)J_{1}(\omega)}{1+\exp(u\omega/4)},
 \label{eq:Egs}
 \\
 \frac{v_{\rms}}{v_{\rmf}}
 &=\frac{I_1(2\pi/u)}{I_0(2\pi/u)},
 \label{eq:vs}
 \\
 \frac{v_{\rmc}}{v_{\rmf}}
 &=\frac{\sqrt{
    \frac{\Delta_{\rmc}}{2t}
    \left[
     1-2\int_{0}^{\infty}\!\! d\omega\,
        \frac{\omega J_{1}(\omega)}{1+\exp(u\omega/2)}
    \right]
   }}
   {1-2\int_{0}^{\infty}\!\!d\omega\,\frac{J_0(\omega)}{1+\exp(u\omega/2)}},
 \label{eq:vc}
 \\
 \frac{\Delta_{\rmc}}{t}
 &=\frac{u}{2}-2
   +4\int_{0}^{\infty}\!\!\frac{d\omega}{\omega}\,
     \frac{J_{1}(\omega)}{1+\exp(u\omega/2)},
 \label{eq:cgap}
\end{align}
where $J_{0}(\omega)$ and $J_{1}(\omega)$ are the Bessel functions, and
$I_{0}(\omega)$ and $I_{1}(\omega)$ are the modified Bessel functions.
We have used the parametrization $u=U/t$, $v_{\rmf}=2at/\hbar$ is the
Fermi velocity in the noninteracting case, and $a$ is the lattice constant.

The effective model (\ref{eq:eff_H}) is expected to be valid in the following two situations:
For weak and intermediate interaction strength, the charge gap lies within the linear part of the spin-excitation band. The model then remains valid for temperatures below or comparable to the charge gap.
For strong interaction strength, the charge gap is larger than the spinon bandwidth. For temperatures well below the energy cutoff given by the spinon bandwidth,
the system is dominated by the spin sector, and well described by the
Heisenberg model or alternatively the linear part of the spin-excitation band.

The quadratic form of the Hamiltonians (\ref{eq:Hs}) and (\ref{eq:Hc})
allows us to immediately compute the partition function and the free
energy $F=E_{\rmgs}+F_{\rms}+F_{\rmc}$ which reads
\begin{align}
 F_{\rms}
 &= -L\frac{\pi}{6\hbar v_{\rms}}(k_{\rmb}T)^2,
 \label{eq:Fs}
 \\
 F_{\rmc}
 &= -L\frac{2k_{\rmb}T}{\pi\hbar v_{\rms}}
    \int_{\Delta_{\rmc}}^{\infty}\!\!d\epsilon\,
    \frac{\epsilon}{\sqrt{\epsilon^2-\Delta_{\rmc}^2}}
    \log\left[1+e^{-\frac{\epsilon}{k_{\rmb}T}}\right],
 \label{eq:Fc}
\end{align}
where $L$ is the system length.
For the Hubbard model, the double occupancy is the derivative of the free energy with respect to interaction strength
\begin{equation}
 n_{\rmd}(T)
 =\frac{\partial F}{\partial U}.
 \label{eq:nd}
\end{equation}
At zero temperature, only the ground state contributes to the double occupancy, i.e.,
$n_{\rmd}(T=0)=\partial E_{\rmgs}/\partial U$,
which is evaluated exactly.
The other contributions determine the finite temperature behavior of the
double occupancy.

\begin{figure}[bp]
 \centering
 \includegraphics[width=\plotwi]{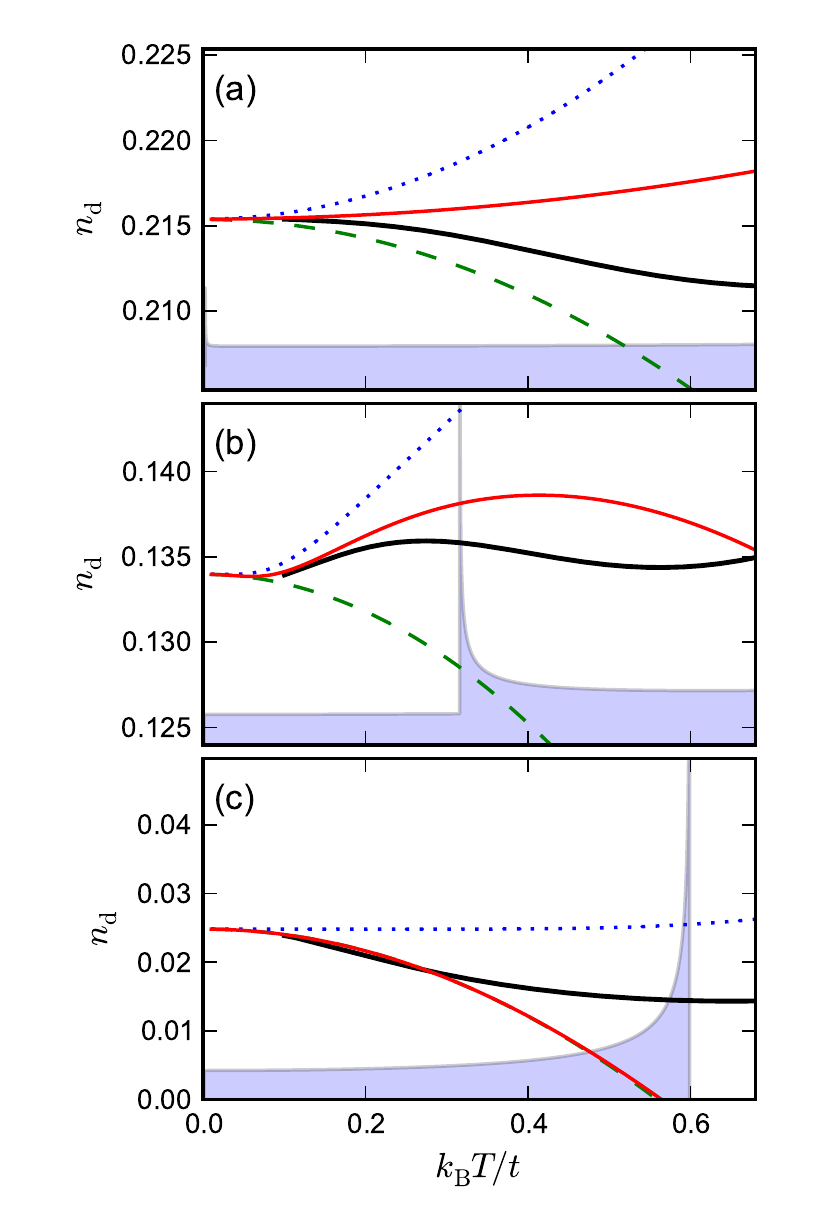}
 \caption{(Color online).
 The double occupancy as a function of temperature: (a) for $U/t=1$,
 (b) for $U/t=3$, and (c) for $U/t=10$.
 Thick red line: double occupancy from the phenomenological argument. The dashed and dotted lines denote the contribution of the spin and charge sectors, respectively.
 Thick black line: DMRG, same data as Fig.~\ref{fig:double_occ} but centered on the low-temperature regime.
 The density of states (up to an arbitrary factor) is also
 shown as a shaded area chart, which is estimated
 by the spinon, particle and hole dispersion given by Bethe
 Ansatz~\cite{Usuki1990}. 
 For $U/t=10$, the contribution from the charge sector is
 negligible, and the temperature dependence of $n_{d}$ is very well described by that of the spin sector in this temperature regime.
 \label{fig:nd-phenomenology}}
\end{figure}

Combining the expression for the double occupancy~(\ref{eq:nd}) with
Eqs.~(\ref{eq:Fs}), (\ref{eq:Fc}), and (\ref{eq:Egs})-(\ref{eq:cgap}),
the temperature dependence of the double occupancy can be computed
straightforwardly. The result for different values of $U/t$ are shown in
Fig.~\ref{fig:nd-phenomenology} together with the corresponding DMRG
result.

As can be seen in Fig.~\ref{fig:nd-phenomenology}, the results for the double occupancy given by the
phenomenological model fit well to the DMRG result as temperature
approaches zero.
In order to analyze the behavior of the double occupancy, we show as well the separate contributions of the spin and charge sector and the density of states. One sees that the charge and spin
sectors always increase and decrease the double occupancy,
respectively, with increasing temperature. The increase caused by the
charge sector is expected, since charge excitations are connected to
particle fluctuations which cause the increase of the double
occupancy. In contrast, the excitation of the spin sector decreases
particle fluctuations. Since at temperatures much below the charge gap,
the gapless spin excitations dominate, the double occupancy is found to
start universally decreasing. This conclusion is in full agreement with the
{\it Pomeranchuk} scenario discussed previously~\cite{Gorelik.etal/PRA85R.2012,Gorelik.etal/PRL105.2010,RichardsonRevModPhys.69.683,DeLeo.etal/PRA83.2011,Dare.etal/PRB76.064402,Werner.etal/PRL95.2005,Georges.etal/PRB48.7167,Paiva/104.066406,Li/85.023624}

At higher temperature the variation of the double occupancy is determined by the competition of both sectors, and as a result can be non-monotonic.
At intermediate interaction strength (see
Fig.~\ref{fig:nd-phenomenology}(b)), the decrease is quickly followed by
a clear rise for $k_BT>0.1t$ due to the onset of charge
excitations. At even higher temperatures the density of states of the spin
sector rises considerably, such that again spin excitations dominate which lead to another decrease of the the double occupancy.
This explains the previously mentioned {\it double Pomeranchuk effect} 
which thus directly stems from the separation of charge and spin excitations for one dimensional systems.

As explained above, our phenomenological model is valid in the temperature regime where the spectrum of spin and charge degrees of freedom is approximately linear and massive, respectively. In Fig.~\ref{fig:nd-phenomenology}(c), one can see the end of the spinon band, signaled by a singularity in the density of states. The phenomenological model holds as expected in the regime of linear spin dispersion, where the density of states is approximately constant.

The observation of the temperature
dependence of the double occupancy is thus intimately
linked to the interplay of
spin and charge modes, which goes beyond the simple Pomeranchuk scenario (decrease and increase only).

As we will see in the next section, the spin charge separation is less
evident in the spin-spin correlation functions, because both types of fluctuations lead to a suppression of correlations.

\subsection{Singlet/Triplet correlations}
\label{sec:singlet_triplet_T}
Recent experimental advances open the possibility to measure the nearest-neighbor singlet and triplet correlations~\cite{Trotzky2008,Greif.etal/Science340.2013}.
This motivates us to study their property in detail. In the sector $S^z =0$, they are defined as:
\begin{align}
&\hat{P}_s(i) = |s\rangle \langle s| \qquad \hat{P}_{t0}(i) = |t_0\rangle \langle t_0| \\
&|s\rangle = \left( |\uparrow_i \downarrow_{i+1} \rangle - |\downarrow_i\uparrow_{i+1} \rangle \right)/ \sqrt{2}\\
&|t_0 \rangle =  \left( |\uparrow_i \downarrow_{i+1} \rangle + |\downarrow_i\uparrow_{i+1} \rangle \right)/ \sqrt{2}
\end{align}
where $|\!\uparrow_i \rangle$ ($|\!\downarrow_i \rangle$) denotes the state with one single atom on site $i$ with spin up (down).
Using the conventional definition of spin operators, $\vec{S}_i = \sum_{\alpha,\beta=\{\uparrow,\downarrow\}} \hat{c}^\dagger_{i \alpha} \vec{\sigma}_{\alpha \beta} \hat{c}^\dagger_{i \beta}$ where $\vec{\sigma}$ denotes the three Pauli matrices, these projectors read $\hat{P}_s(i) = 1/4 \hat{P}^1_i \hat{P}^1_{i+1} - {\vec{S}}_i \cdot {\vec{S}}_{i+1}$, $\hat{P}_{t0}(i) = 1/4 \hat{P}^1_i \hat{P}^1_{i+1} + {\vec{S}}_i \cdot {\vec{S}}_{i+1} - 2  \hat{S}^z_i \hat{S}^z_{i+1}$. Here $\hat{P}^1_i$ is the local projector onto singly occupied states on site $i$.
As a consequence, singlet and triplet correlations depend on both spin and charge correlations. Due to the spin rotation $\mathrm{SU}(2)$ symmetry, one can simplify the above relations using $\langle {\vec{S}}_i \cdot {\vec{S}}_{i+1} \rangle = 3 \langle  \hat{S}^z_i \hat{S}^z_{i+1} \rangle$.

\begin{figure}
\includegraphics[width=\plotwi]{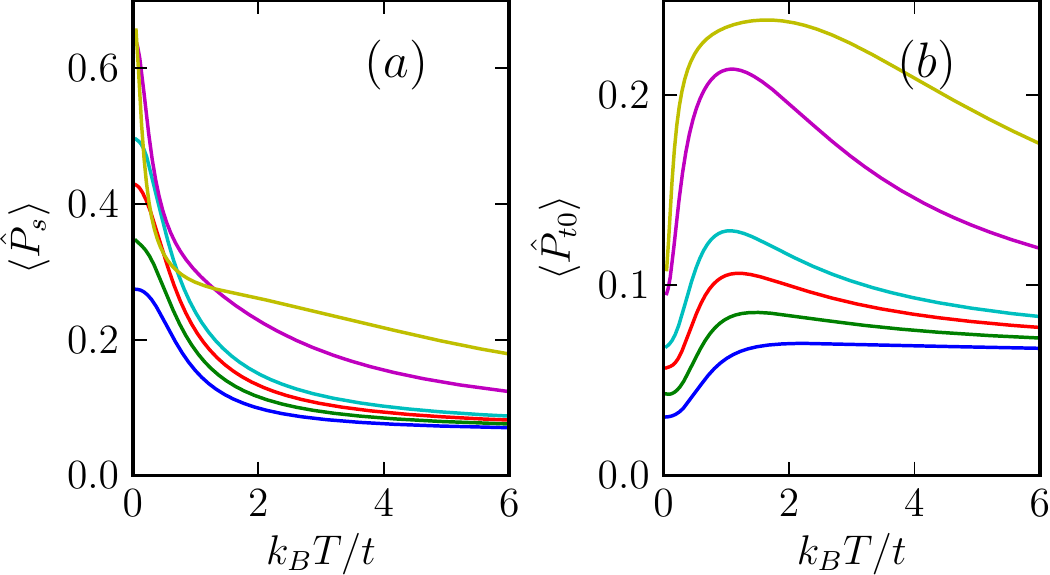}
\caption{\label{fig:singlet_triplet_T}(Color online). Spin correlations in a homogeneous chain at half-filling for $U/t = 1,2,3,5,10,20$
from bottom to top increasing $U/t$ (high $k_BT/t$). (a) Singlet correlations. The number of singlets is maximal in the ground state and decays monotonically with growing $T$. (b) Triplet $S^z =0$ correlations. At low temperature triplet excitations are created, and they decay again at higher temperatures.}
\end{figure}

At low temperatures, the singlet correlations are higher than triplet correlations, as shown in Fig.~\ref{fig:singlet_triplet_T}. The ground state singlet correlations increase with $U/t$, due to the suppression of charge fluctuations.
As temperature increases, more and more triplet excitations are created and the singlet correlations decrease. The corresponding energy scale is of the order of the tunneling $t$ at low $U/t$, and is the antiferromagnetic exchange $\Jaf$ at high $U/t$.
In contrast, the triplet correlations are very small at low temperatures and then increase with rising temperature, since excitations in the spin sector are created and dominant. At higher temperatures, the triplet correlations decay due to creation of charge fluctuations, seen in the increase of the doubly occupied states.

We have thus a good characterization of the degree of short range antiferromagnetic order as a function of temperature. 
An interesting question is how the quasi-long range order (for one dimension no long range
antiferromagnetic order exists) is connected to the short range one. In order to explore this issue 
we investigate the decay of the correlations with distance, since in one dimension at finite temperature correlation functions decay exponentially \cite{giamarchi_book_1d}. 

\begin{figure}
\includegraphics[width=\plotwi]{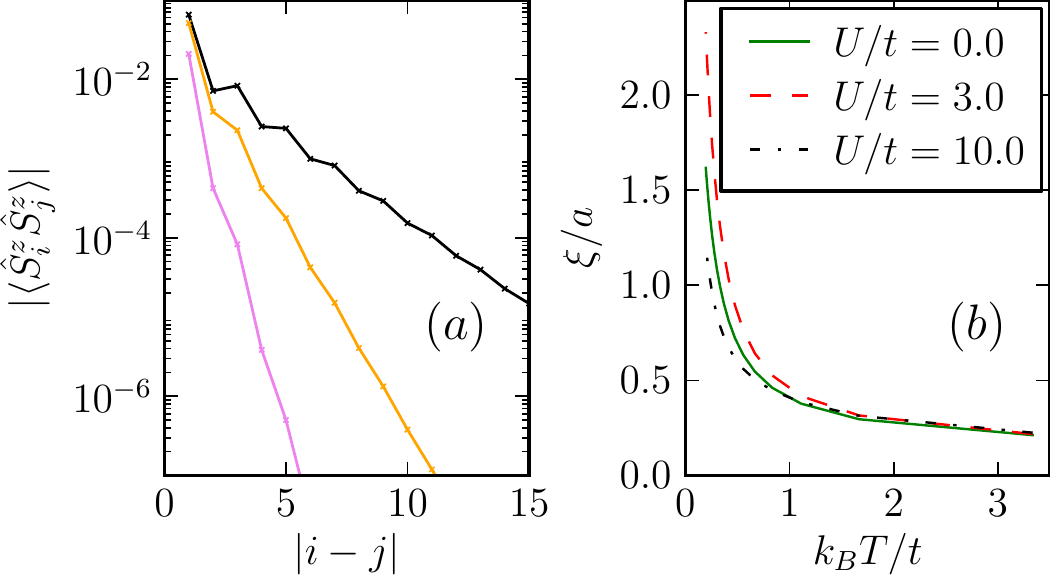}
\caption{\label{fig:correlation_length}(Color online). (a) Spin correlations $\langle \hat{S}^z_i \hat{S}^z_{i+1} \rangle$ as a function of distance $|i-j|$, log-lin scale, $U/t=1.5$, $k_B T/t=0.20,0.48,1.11$ from top to bottom increasing $T$.
The correlations decay exponentially for large distances and increase while lowering $T$.
The DMRG relative error for distances $|i-j|\lesssim 12$ is below one percent comparing the number of retained states $M=192$ and $M=512$.
(b) Correlation length $\xi$ as a function of temperature $k_B T/t$ for $U/t = 1.5,5,10$ from top to bottom (low $T$).
The value of $\xi$ represented here is obtained with a number of retained states $M=512$.
 The deviations with respect to the number of states $M=192$ and $M=512$ is below one percent, see Fig.~\ref{fig:correlation_lengthb} for the convergence with respect to the fit range.
}
\end{figure}

Typical results for the spin-spin correlations as a function of distance are shown in Fig.~\ref{fig:correlation_length}.

\begin{figure}
\includegraphics[width=\plotwiless]{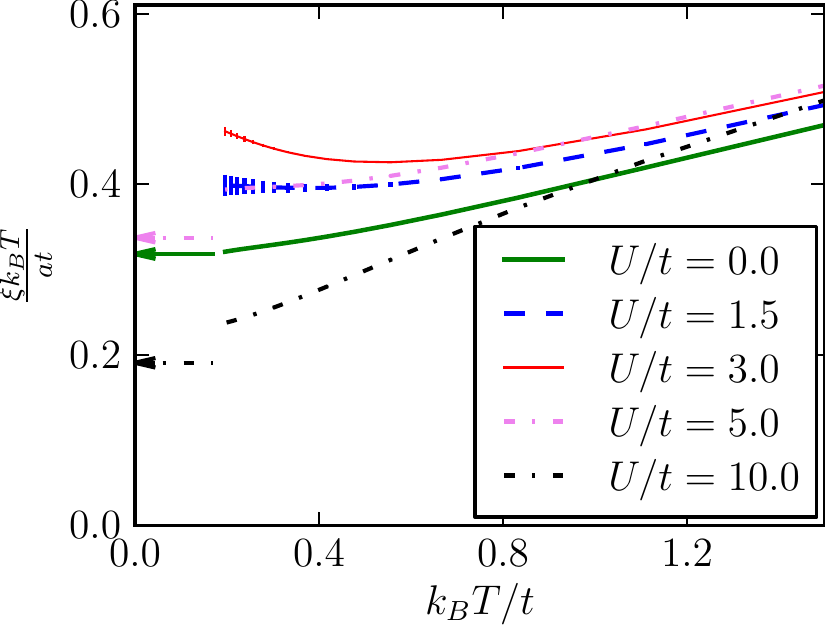}
\caption{\label{fig:correlation_lengthb}(Color online). Rescaled spin correlation length $\xi k_B T/at$ versus temperature. In the Tomonaga Luttinger liquid regime, the conformal field theory predicts that $\xi k_B T/at$ goes to a constant for low $T$. We compare it with the prediction of the Tomonaga Luttinger liquid for $U/t= 5,0,10$ in this order from top to bottom (arrows). The correlation length is estimated on a range $|i-j|$ in $[10,20]$ and the error bar is the difference with the estimate for a range $[6,12]$, the small ranges being the main source of error.
}
\end{figure}

A full analysis of the long range behavior would go beyond the scope of the present paper, but some elements of information can be readily extracted. 
From the field theory of one dimensional systems \cite{giamarchi_book_1d} one can expect the spin-spin correlation to behave 
asymptotically as
\begin{equation} 
\langle S_r \cdot S_0 \rangle = A(r) + (-1)^r B(r)
\end{equation} 
where $A(r)$ is the ferromagnetic component, describing fluctuations with momentum close to zero and $B(r)$ the antiferromagnetic one 
describing spin fluctuations with momentum close to $\pi$. In the low energy limit $A(r)$ only depends on the spin part of the Hamiltonian 
and thus on $v_\sigma$, while $B(r)$ depends both on the charge and the spin parts. At zero temperature $A(r)$ decays as $1/r^2$ while 
$B(r)$ decays with a non-universal power law (for example $1/r$ for large $U$). At finite temperature, as can be shown by conformal invariance 
a power law turns into an exponential decay~\cite{giamarchi_book_1d} with a correlation length of the form 
\begin{equation} \label{eq:conformal}
 \xi_\nu = (\hbar \beta v)/( \nu \pi)
\end{equation}
where $\beta = 1/(k_B T)$ is the inverse temperature, $\nu$ the zero temperature exponent, and $v$ the velocity of the corresponding
excitations. We thus see that in general the spin-spin correlation will be characterized by two correlation lengths. However for the 
parameters considered here we find that the part $A$ decays very rapidly and the correlations are dominated by the oscillating part $B$.
Thus we extract the correlation length from a single exponential fit as shown 
in Fig.~\ref{fig:correlation_length}(b) for various temperatures. We see a clear divergence at low temperature as can be expected 
from (\ref{eq:conformal}). The general temperature dependence of the correlation length is shown in Fig.~\ref{fig:correlation_lengthb}.

The field theory formula (\ref{eq:conformal}) would predict a saturation of $\xi k_B T$ at low temperature as roughly observed in the numerical data. For very small or very large $U$ one can quantitatively compute the correlation lengths as 
$\xi/a=t/(\pi k_{B}T)$ for $U/t=0$ and
$\xi/a=\hbar v_{s}/(a \pi k_{B}T)$ in the large $U/t$ limit, with $v_s$ defined in Eq.~\eqref{eq:vs}.

These limiting values (indicated with arrows in Fig.~\ref{fig:correlation_lengthb})
agree reasonably well with the extrapolation of the numerical data. For intermediate $U$ values and larger temperatures there are of course
many corrections to the asymptotic low energy formulas coming in particular from the coupling of the spin and charge degrees of freedom at intermediate energies. 

We conclude this section with a few observations. First, the value $\xi/a \gtrsim 2$ is reached at a fraction of the temperature on which the nearest-neighbor correlations grow.
The correlation length also grows faster for intermediate $U$ than in the limit of large and small $U$. This is in qualitative agreement with the fact that the nearest-neighbor correlations are the largest at intermediate $U$ (see Fig.~\ref{fig:colorplota}) and suggests that 
the measure of the nearest neighbor correlations is indeed a good way to detect the onset of quasi-long range antiferromagnetic order.

\section{Trapped system}
\label{sec:trap}
Since in current experimental realizations a trapping potential is present, we would like to understand its effect on the singlet and triplet correlations.
We focus in particular on the parameters used in Ref.~\cite{Greif.etal/Science340.2013}.

\subsection{Singlet and triplet correlations in the trap}
\label{sec:trap_singlet_triplet}
In the following, the system is described by the Hamiltonian of Eq.~(\ref{eq:hamiltonian}) with an additional harmonic potential $V_i = V_t (i-L/2)^2$ which is coupling to the local density.

We find that due to thermal redistribution of the particles within the trap, the dependence of averaged spin correlations on temperature can differ from the case of a homogeneous system discussed above (Fig.~\ref{fig:singlet_triplet_T}).

We discuss two different density regimes, with a density close to $n=1$ and
$n=2$ at the center of the trap at low temperatures $k_B T/t \lesssim 1$.
The density profile can be better characterized by the rescaled densities $\rho = \left( \frac{V_t}{zt} \right)^{d/\alpha} N$, with $\rho=1.61,4.82$ here~(see also Appendix~\ref{sec:rescaled_density}).

\begin{figure}

\includegraphics[width=\plotwi]{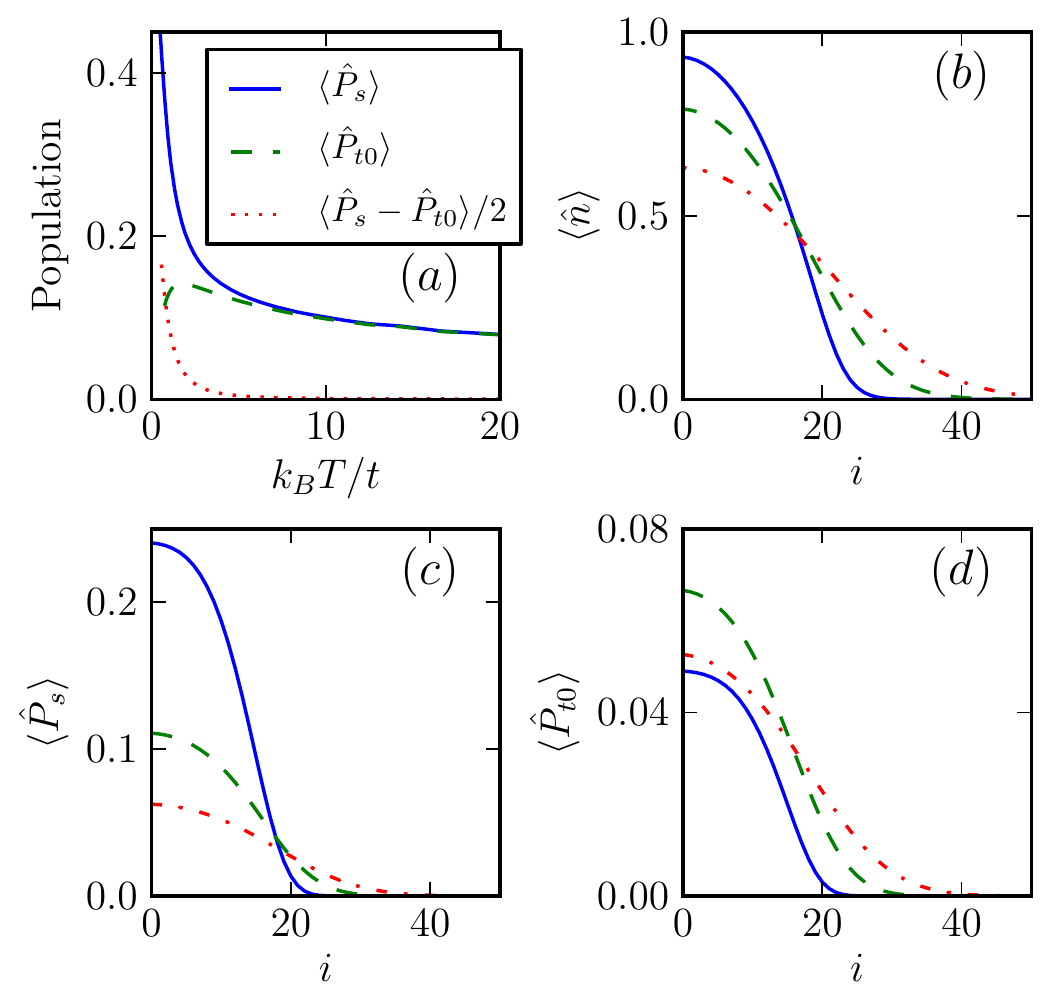}
\caption{\label{fig:schain30}(Color online). (a) Singlet and triplet correlations as a
 function of $k_BT/t$ averaged over the chain in the presence of a trapping potential, for a rescaled
 density of $\rho=1.61$ and $U/t =1.44$. (b) Density distribution, (c) local singlet
 and (d) triplet correlations along the chain, $k_BT/t=0.5,1.5,3$ for the solid, dotted and dash-dotted lines, respectively.}
\end{figure}

For the low rescaled density $\rho=1.61$ (Fig.~\ref{fig:schain30}(a)), at first sight, the behavior of the singlet and triplet correlations resembles the one of the homogeneous system. In particular, the singlet correlations decay monotonically with increasing temperature and the triplet correlations show a maximum around $k_B T/t=1.55$. However, due to the redistribution of the density by thermal excitations (Fig.~\ref{fig:schain30}(b)), the correlations decrease more rapidly than they would in a homogeneous system.
This is due to the fact that doping away of half filling reduces the spin correlations as seen in Fig.~\ref{fig:schain30}(c) and~\ref{fig:schain30}(d).

\begin{figure}

\includegraphics[width=\plotwi]{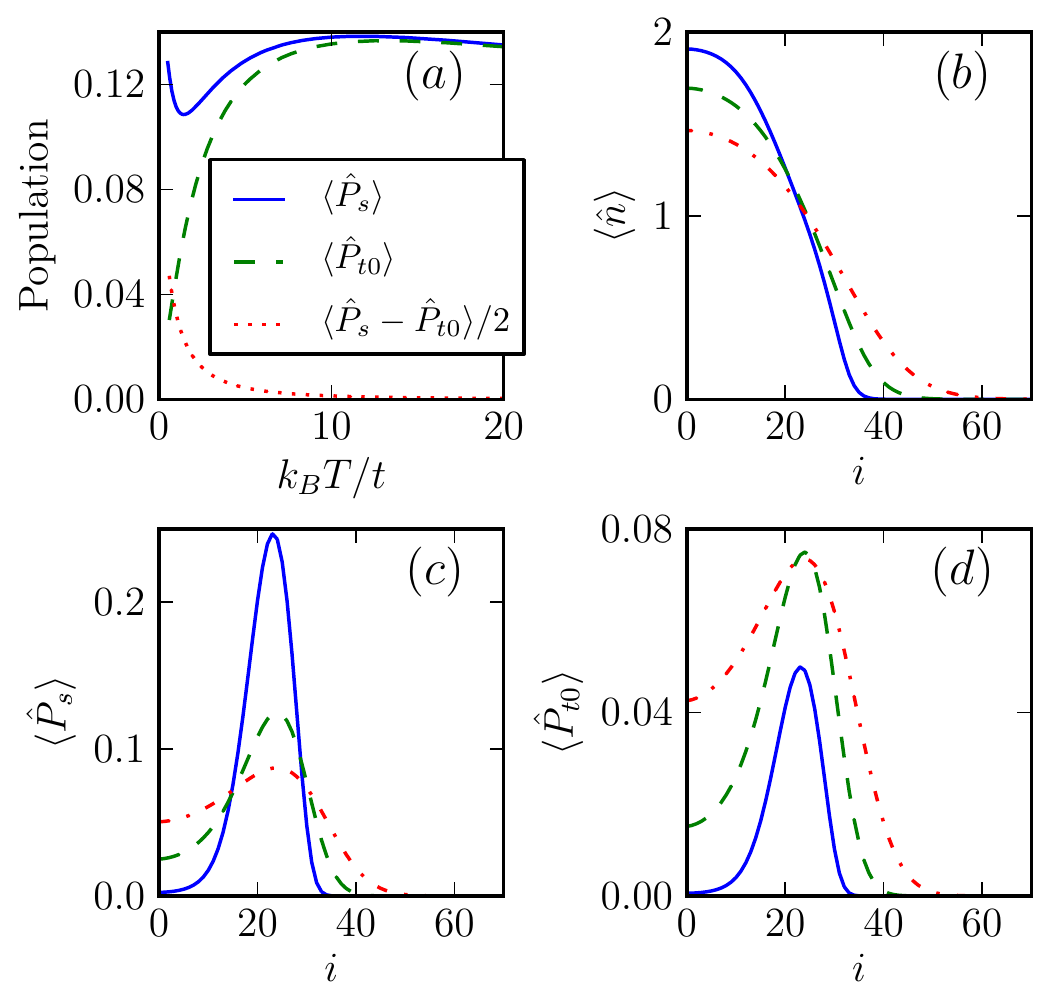}
\caption{\label{fig:schain90}(Color online). (a) Singlet and triplet correlations as a
 function of $k_BT/t$ averaged over a single trapped chain, for a rescaled
 density of $\rho=4.82$ and $U/t =1.44$. (b) Density distribution, (c) local singlet
 and (d) triplet correlations along the chain, $k_BT/t=0.5,1.5,3$ for the solid, dotted and dash-dotted lines respectively.}
\end{figure}

In contrast, for high rescaled density $\rho=4.82$, the singlet correlations behave very differently and show a non-monotonic temperature dependence (Fig.~\ref{fig:schain90}(a)).
This non-monotonicity can be explained by the competition between the intrinsic decrease which is counteracted by the particle redistribution.
At very low temperatures the intrinsic decrease of the singlet correlations is dominating. This first decrease is followed by a strong increase of correlations. As seen in Fig.~\ref{fig:schain90}(b), the particles are redistributed from the center towards the edges for increasing temperature. This leads in the central part of the trap to a depletion of the local density towards half filling $\langle{\hat{n}}\rangle=1$.
Since the singlet correlations are maximized at half-filling (see maximum around $i=22$ in Fig.~\ref{fig:schain90}(c)), this leads to an increase of the local singlet correlations in the central region of the trap (Fig.~\ref{fig:schain90}(c)). This increase in the central region overwhelms the intrinsic decrease which occurs in the boundary regions at intermediate temperatures.
For even larger temperatures the redistribution is less relevant and the intrinsic decay dominates. Thus, the competition between the intrinsic singlet correlation decay and the enhancement of correlations by particle
redistribution is the cause of this non-monotonicity.

The non-monotonic temperature dependence of the singlet and triplet
correlations found makes their use as a thermometer difficult.
However, for the parameters considered here the difference of the two
is monotonic as seen in Figs.~\ref{fig:schain30}(a) and \ref{fig:schain90}(a), which can be a better measure of temperature. Indeed, the difference of singlet and triplet correlations is a function of magnetic correlations only, which have monotonous temperature variations.

\subsection{Local observables in the trap}
\label{sec:trap_lda}
\begin{figure}
\includegraphics[width=\plotwi]{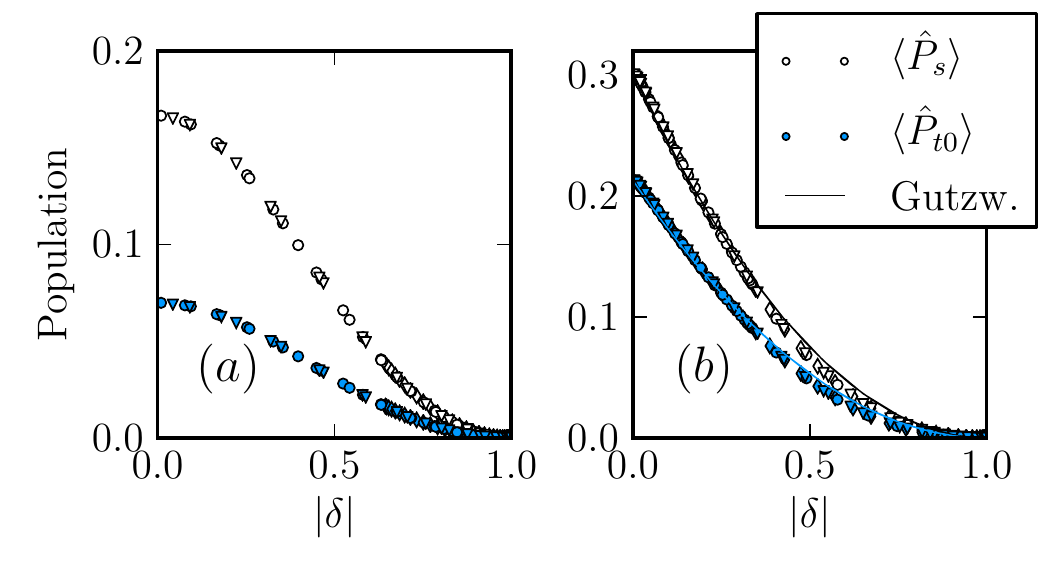}
\caption{\label{fig:lda}(Color online). Singlet and triplet correlations
as a function of the local doping $\delta = 1- \langle \hat{n}
\rangle$, collapsed data for several trapped chains with different
rescaled densities $0<\rho<5$ (various markers), $k_BT/t=1$. (a) $U/t=1.5$, (b) $U/t=10$.
The Gutzwiller approximation for $U/t=10$ is also depicted as lines, see text.}
\end{figure}

In order to check the quality of the local density approximation we plot the singlet and triplet correlations for different characteristic densities $0<\rho<5$ at $k_BT/t=1$ versus the local doping $\delta =  1- \langle \hat{n} \rangle$ in Fig.~\ref{fig:lda}. Since all the data collapses on a single line,
this indicates that the actual value of the correlations is fully determined by the local density.
We also checked that the local density versus local chemical potential dependence collapses to a good precision, indicating that the local density approximation holds.

The spin correlations are maximal at half-filling since
empty or doubly occupied sites do not contribute to spin correlations.
The decay of the spin correlations is found to become sharper for larger
interaction values $U/t$. It is explained as follows: for strong interactions, the
doping systematically induces double occupancies, which in turn suppress the spin
correlations. Therefore, the dependence on the doping around half filling is approximately linear.
Using the Gutzwiller approximation~\cite{gebhard/1997} one obtains in the limit of large $U/t$:
$P_{s/t0}(\delta) = P_{s/t0}(\delta=0) \times (1-|\delta|)^2$. This result is shown as lines in Fig.~\ref{fig:lda}, and is in good agreement with
the numerical results for $U/t=10$ for low doping.

On the other hand, for weaker interactions, the density fluctuations
are already non negligible at half-filling, and the additional particles
are partially redistributed on empty sites, limiting the formation of double occupancies, and thus making the suppression of spin correlations by the doping softer.

\section{Comparison to experimental results}
\label{sec:trap_exp}
In current experiments, the one-dimensional tubes are obtained using a very strongly anisotropic three-dimensional lattice as in Ref.~\cite{Greif.etal/Science340.2013}, with a strong tunneling $t$ along the $x$-axis and weaker tunneling $t_\bot$ along the transverse $y$ and $z$ axis.
In this section, we consider the array of trapped one-dimensional tubes, neglecting the weaker coupling $t_\bot$ between chains.
This approximation is typically valid for temperatures $k_BT \gtrsim t_\bot$.
The trapping potential is assumed to be harmonic
$V_{ijk}= V_t(i^2+\alpha_y j^2+\alpha_z k^2)$ where $(i,j,k)$ are
indices along the $x,y,z$ axis, with $V_t/t = 5.75 \times 10^{-3}$,
$\alpha_y = 2.84$, and $\alpha_z = 0.84$.
The interaction strength is $U/t =1.44$, and the global chemical potential is set to reproduce the
total number of atoms $N=66000$ for each temperature. For these parameters, the interactions are too small to form clear Mott plateaus.

In the experiment~\cite{Greif.etal/Science340.2013}, the density difference
$\langle \hat{P}_s -\hat{P}_{t0} \rangle/2$ has been
measured for a set of different initial entropies per particle $S_i$
\emph{before} loading atoms into the lattice.
Here we estimate the temperature $\Tf$ and entropy per
particle $\Sf$ after the atoms have been loaded into the lattice. They are
determined such that the simulated spin correlation as a function of
$\Tf$ (or of $\Sf$) matches the experimentally obtained value.
The correspondence is unique for the cases we considered, as was the case
for single chains in Figs.~\ref{fig:schain30}(a) and
\ref{fig:schain90}.
Note that we neglect altogether the small amount of entropy along the transverse
directions $y,z$, since this calculation is based on the simulation of isolated chains.

\begin{figure}
\includegraphics[width=\plotwi]{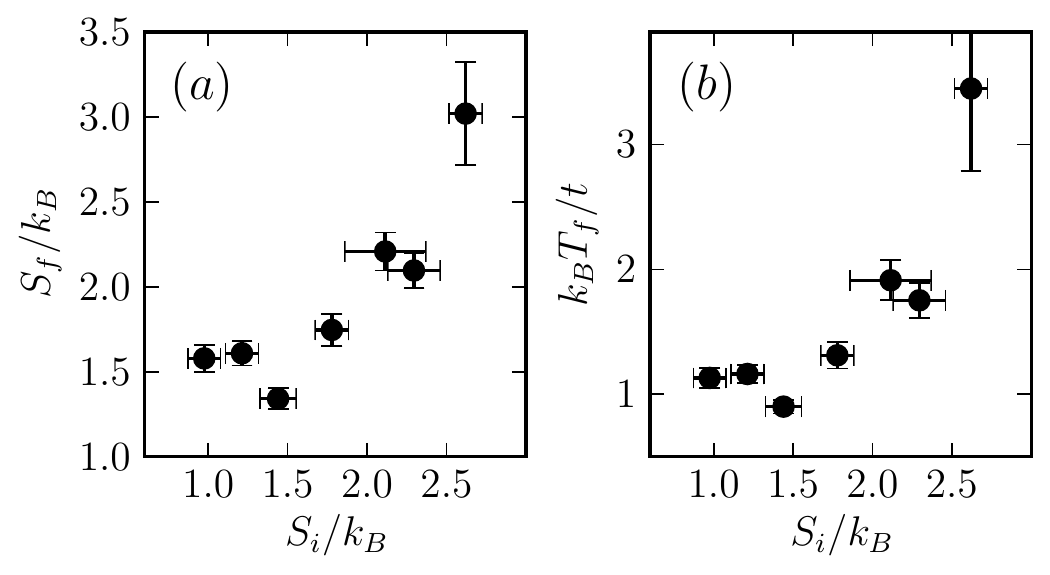}
\caption{\label{fig:temp_entropy_determ}(Color online). The final
 entropy $\Sf$ (a) and temperature $k_B \Tf/t$ (b)
 for the experiment~\cite{Greif.etal/Science340.2013}. Both are deduced from the difference
 of the singlet and triplet correlations as a function of the
 experimental initial entropy $S_i$. The error bars
 on $S_i$ have been estimated in~\cite{Greif.etal/Science340.2013}. The
 uncertainty on $\Sf$ and $k_B \Tf/t$ is the propagated uncertainty of
 $\langle \hat{P}_s-\hat{P}_{t0} \rangle/2$
 from~\cite{Greif.etal/Science340.2013}, the DMRG errors being much
 smaller. We also take into account the uncertainty due to the
 fluctuations of the number of atoms $N=66000 \pm 6000$ and they are
 found to add very little to the estimated uncertainty.
 Experimental data courtesy
 of Greif et al.~\cite{Greif.etal/Science340.2013}. 
}
\end{figure}

The results for $\Sf$ and $\Tf$ are shown in Fig.~\ref{fig:temp_entropy_determ}.
For large initial entropy $S_i>1.5k_B$ the dependence of the final entropies on the initial one is almost linear. Typically the determined entropy $\Sf$ is a bit larger than the initial entropy $S_i$.
This effect is even larger at lower initial entropy, where the determined entropy $\Sf$ approximately saturates.
In this regime, the final entropy $\Sf$ in the optical lattice is approximately $0.4k_B$ larger than the initial entropy $S_i$, which might stem from the heating during the lattice loading procedure, a value consistent with the previous estimation~\cite{Greif.etal/Science340.2013}.
For larger initial entropies, the heating seems less pronounced.
We also extract the corresponding final temperature (Fig.~\ref{fig:temp_entropy_determ}(b)), which has a very similar behavior.
From our data, we see that the lowest temperatures reached in the experiment~\cite{Greif.etal/Science340.2013} are $k_{B}\Tf /t \approx 1\pm0.1$, which corresponds to the onset of nearest-neighbor spin correlations as shown in Sec.~\ref{sec:singlet_triplet_T}.

\begin{figure}
\includegraphics[width=\plotwiless]{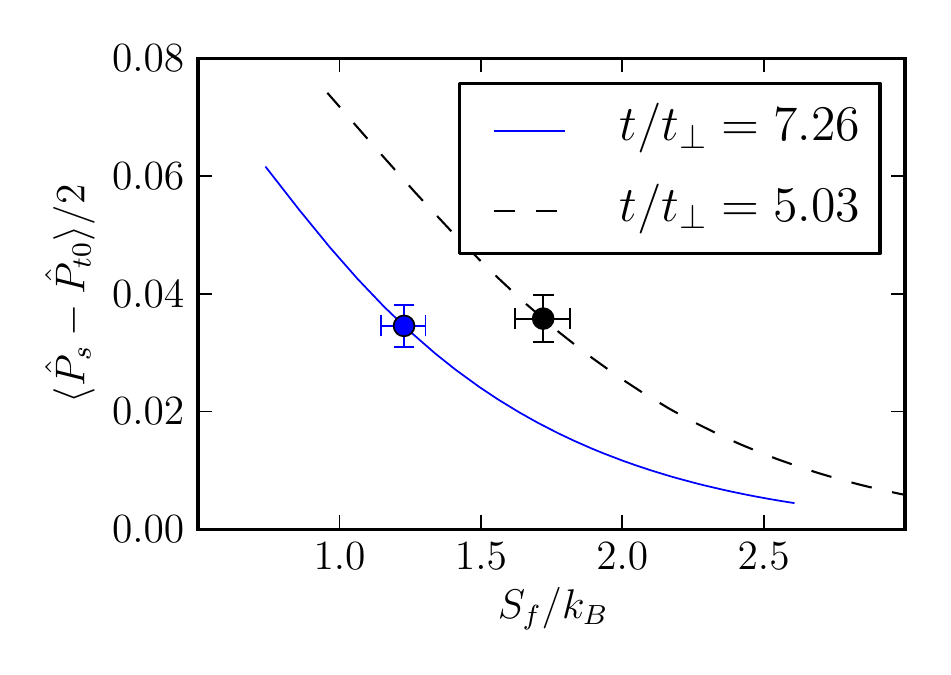}
\caption{\label{fig:entr_4a} (Color online). The singlet-triplet
 population imbalance as a function of the final entropy $S_f$ for
 $t/t_\bot = {7.26,5.03}$. The dots denote experimental singlet-triplet
 population imbalance from~\cite{Greif.etal/Science340.2013} and the
 deduced final entropy. The error bars on the population are from the
 experimental ones~\cite{Greif.etal/Science340.2013} and the propagated
 error on the entropy is represented. All other uncertainties are neglected (see Fig.~\ref{fig:temp_entropy_determ} for details).}
\end{figure}

We also compute the entropy and temperature for two different tunnelings
$t$ corresponding to $t/t_\bot = 7.3$ and $5.0$ in Fig.~4(a) of
Ref.~\cite{Greif.etal/Science340.2013}, with interaction strength
$U/t =1.43$ and $2.36$ and trapping potential $V_t/t=5.5\times10^{-3}$ and $8.0\times10^{-3}$, respectively~\footnote{The transverse trapping is $\alpha_y = 2.29,3.51$ and $\alpha_z=0.82,0.97$ for $t/t_\bot = 7.3,5.0$, respectively.}. As shown in Fig.~\ref{fig:entr_4a}, although the experimental population imbalance in singlet-triplet correlations have the same value $\langle \hat{P}_s-\hat{P}_{t0} \rangle/2 \eqsim 0.035$ (within error bars) for the two values of $t$, the deduced entropy $S$ is actually larger by a factor of 1.4 for the weaker $t$.
This is consistent with our previous results, since a larger temperature
(larger entropy) is needed to get similar correlations for larger $U/t$
in the regime $U/t\lesssim 3$. (see Fig.~\ref{fig:singlet_triplet_T}.)
One of the possible reasons for the entropy to be larger in the case of a lower tunneling $t$, may be that the loading procedure induces more heating since the final lattice potential is higher.

\section{Conclusion}
\label{sec:conclusion}
In this paper, we have studied various aspects of the thermodynamics of
the one-dimensional Hubbard model, both in the infinite length limit and in trapped setups.
In the first part, we have described in detail the low-temperature properties of the double occupancy. The Pomeranchuk effect occurs in its most simple form at large $U/t$, where the double occupancy $n_d(T)$ first decreases and then increases with increasing temperature.
For intermediate interactions, however, the non-monotonicity is more complex.
To understand the underlying physics, we have considered an effective theory including both
the spin degrees of freedom with linear dispersion and charge
degrees of freedom with gaped dispersion.
This treatment yields a clear picture of the Pomeranchuk effect in terms
of the low-lying excitation modes, in particular, in the non-trivial intermediate $U/t$ regime.

In the second part, we obtained the low-temperature singlet and triplet
nearest-neighbor correlations and their dependence on interactions. From
the study of longer range correlations, we observe that the temperature
of the onset of short-range correlations is slightly above the temperature
of growth of correlation length.
Below this temperature, we find a fair agreement of the correlation
length with the predictions of conformal field theory 
in the regime of large interactions.

Additionally, we have discussed the effect of a trap potential onto the temperature dependence of correlations.
We remark that the density redistribution effect can play a significant role, and can induce a non-monotonic dependence of singlet correlations with temperature, depending on the density regime.
We also present an extensive comparison with the experiment~\cite{Greif.etal/Science340.2013} to extract the temperature and entropy in various setups.
We estimate that $k_{B}T/t \approx 1\pm0.1$ is reached in the experiment, close to the temperature where longer distance correlations are established.

\section{Acknowledgments}
We thank the group of T.~Esslinger for fruitful discussions and for providing the experimental data. Our computer simulations employed the ALPS libraries \cite{Albuquerque2007,Bauer2011}.
We acknowledge ANR (FAMOUS), SNSF under Division II and MaNEP for their financial support.
TG would like to thank the Aspen Center for Physics and the NSF Grant 1066293 for hospitality during the writing of this paper.

\appendix
\section{Rescaled density}

\label{sec:rescaled_density}
For completeness in this appendix, we discuss the properties of the reduced density $\rho$, following the notation of Ref.~\cite{DeLeo.etal/PRA83.2011}. This was first introduced in the context of one-dimensional Bose-systems \cite{Rigol.etal/PRL91.130403}.
For the sake of generality, consider a $d$-dimensional lattice of lattice spacing $a$, within a spherically symmetric potential $V(\mathbf{r})=V_t |\mathbf{r}/a|^\alpha$.

Within the local density approximation, the induced local chemical potential is $\mu(\mathbf{r}) = \mu_0 - V_t |\mathbf{r}/a|^\alpha$ and the average of any local observable $\langle \hat{\mathcal{O}}(\mathbf{r}) \rangle$ is uniquely determined by its homogeneous counterpart evaluated at the corresponding chemical potential $\mathcal{O}(\mu(\mathbf{r}) )$. Here $\mu_0$ denotes the chemical potential in the center of the trap. In this situation the average over the trap of a local operator $\hat{\mathcal{O}}(\mathbf{r})$ reads
\begin{align}
\frac{\overline{\mathcal{O}}}{N} &= \frac{\Omega_{d-1}}{a^d} \int \rmd {\bf r} \; r^{d-1} \mathcal{O} (r) \nonumber\\
 &= \frac{1}{\rho} \frac{\Omega_{d-1}}{\alpha} \int^{\overline{\mu}_0(\rho)}_{-\infty} \rmd \overline{\mu} (\overline{\mu}_0(\rho)-\overline{\mu})^{d/\alpha-1} \mathcal{O} (\overline{\mu}) \label{eq:rescaled}
\end{align}
In the last expression we changed variables to $\overline{\mu} = \frac{\mu}{zt}$ with $z$ the lattice coordination, and defined
\begin{align}
\rho = \left( \frac{V_t}{zt} \right)^{d/\alpha} N
\end{align}
The corresponding formula for the particle number can be used in order to determine the central chemical $\overline{\mu}_0$ which only depends on the characteristic density. Therefore, given the dimensionality $d$ and on the nature of the trap through the exponent $\alpha$, two systems with the same reduced density have the same integrated properties, although their respective trapping potential $V_t$ and number of particle $N$ may be very different. This quantity is useful to establish generic state diagrams~\cite{DeLeo.etal/PRA83.2011} in trapped setups.

\bibliography{ref.bib}

\end{document}